\documentclass[12pt,preprint]{aastex}

\shorttitle{Star Formation around the Ori-Eri Superbubble}

\shortauthors{H.-T. Lee and W. P. Chen}

\begin{document}

\title{TRIGGERED STAR FORMATION ON THE BORDER OF THE ORION-ERIDANUS SUPERBUBBLE}

\author{Hsu-Tai Lee\altaffilmark{1,2}} 
\email{htlee@asiaa.sinica.edu.tw}

\author{W. P. Chen\altaffilmark{1,3}} 
\email{wchen@astro.ncu.edu.tw}

\altaffiltext{1}{Institute of Astronomy, National Central University, Taiwan 320, R.O.C.}

\altaffiltext{2}{Institute of Astronomy and Astrophysics, Academia Sinica, P.O. Box 23--141, Taipei 10617, Taiwan, R.O.C.}

\altaffiltext{3}{Department of Physics, National Central University, Taiwan, 320, R.O.C.}

\begin{abstract}

A census of classical T Tauri stars and Herbig Ae/Be stars has been performed around the Orion-Eridanus 
Superbubble which is ionized and created by the Ori~OB1 association.  This sample is used to study the 
spatial distribution of newborn stars, hence the recent star formation sequence, in the region that 
includes two giant molecular clouds (Orion~A and B) and additional smaller clouds (NGC~2149, GN~05.51.4, 
VdB~64, the Crossbones, the Northern Filament, LDN\,1551, LDN\,1558, and LDN\,1563).  Most of the molecular clouds  
are located on the border of the Superbubble, and associated with H$\alpha$ filaments and star formation 
activity, except the Northern Filament which is probably located outside the Superbubble.  This suggests 
that while star formation progresses from the oldest Ori\,OB1a subgroup to 1b, 1c and 1d, the Superbubble 
compresses and initiates starbirth in clouds such as NGC\,2149, GN\,05.51.4, VdB\,64, and the Crossbones, 
which are located more than one hundred pc away from the center of the Superbubble, and even in clouds 
some two hundred pc away, i.e., in LDN\,1551, LDN\,1558, and LDN\,1563.  A superbubble appears to have 
potentially a long-range influence in triggering next-generation star formation in an OB association.  

\end{abstract}


\section{INTRODUCTION}

Ori~OB1 is one of the nearest OB associations \citep{dez99}, harboring some 10 O-type stars 
\citep{hum78}.  \citet{bla64} divides the Ori~OB1 association into four subgroups, Ori~OB1a, 1b, 1c, 
and 1d, with 1a being the oldest and 1d (the Orion Nebula) being the youngest \citep{bro94,war78,bla91}.  
From the $Hipparcos$ catalogue, \citet{dez99} determine distances of 336$\pm$16, 473$\pm$33, and 
506$\pm$37 pc to Ori~OB1a, 1b, and 1c, respectively.

In Ori~OB1, most of the current star formation activity is located in, or close to, the two giant 
molecular clouds \citep{wal00,meg05,her07a,bri07a}, Orion~A and B, which have been surveyed in $^{12}$CO 
by \citet{mad86} and \citet{wil05}.  Together with many smaller molecular clouds, the entire region is 
called the Orion-Monoceros complex \citep{mad86,wil05}.

As the Ori~OB1 association evolved, the ionization photons and stellar winds from massive stars 
and supernova ejectra produced the Orion-Eridanus Superbubble \citep{bro95}.  The Orion-Eridanus 
Superbubble has been studied in many wavelengths \citep[reviews by][]{hei99}, e.g., in \ion{H}{1} 
\citep{hei76,bro95}, $IRAS$~100~$\micron$ \citep{bur93}, H$\alpha$ \citep{siv74,rey79}, X rays 
\citep{bur93,sno95,guo95}, and $\gamma$ rays \citep{par98}.

Even though the Orion-Eridanus Superbubble is one of the most explored superbubbles, the overall star 
formation related to the Superbubble is not well investigated, simply because of its large sky coverage.  
It is indeed a great challenge to identify low-mass pre-main-sequence (PMS) stars in such a large region, 
so a wide-field survey is needed.  \citet{wal00} and \citet{bri07a} summarize different methods to 
search for low-mass (0.1--1~M$_{\sun}$) PMS stars in OB associations, e.g., by objective-prisms, X 
rays, proper motions, and variability surveys.  Each method has its own advantage and drawback and here we 
summarize briefly each of the methods that have been applied to the Orion region.

An objective-prism imaging finds emission-line stars as young star candidates.  A large number of 
candidate PMS stars have been recognized by this method, e.g., in Orion \citep{wir89,wir91,wir93,kog89} 
by the Kiso H$\alpha$ survey.  The total number of H$\alpha$ emission stars is 1157 over an area of 150 
square degrees \citep{wir93} to a limiting magnitude of V=15.  However, an H$\alpha$ survey detects  
PMS stars only with strong H$\alpha$ emission, i.e., classical T Tauri stars (CTTSs) and Herbig Ae/Be 
(HAeBe) stars, but is not sensitive to weak-line T Tauri stars (WTTSs).  An object-prism sample may 
also be contaminated considerably by late-type dwarfs that also show H$\alpha$ in emission, namely the 
dMe stars.  Indeed, \citet{bri01} conclude that only about 40\% of the H$\alpha$ emission stars in Orion 
OB1a found in the Kiso H$\alpha$ survey are really PMS stars.

Because low-mass PMS stars (CTTSs and WTTSs) are X-ray sources \citep{fei99}, the $ROSAT$ All Sky 
Survey (RASS), for example, has been used for large-scale searches for low-mass PMS stars.  $ROSAT$ 
is sensitive to soft X rays, so would find preferentially WTTSs.  The main contaminations in an RASS 
sample would be chromospherically active K and M dwarfs.  \citet{ste95} study the density distribution 
of X-ray-selected PMS candidate stars in Ori OB1 (700 square degrees) and find significant surface 
density enhancements around Ori OB1a, OB1b, OB1c, and $\lambda$ Ori.  \citet{alc96} accomplish 
spectroscopic and photometric observations of the RASS X-ray sources in the Orion cloud 
complex and find 112 new WTTSs in about 450 square degrees, and follow up by broad- and narrow-band 
photometric and high-resolution spectroscopic observations \citep{alc98,alc00}.
\citet{wal00} compare the spatial distributions of the H$\alpha$ emission stars in the Kiso 
H$\alpha$ survey and RASS X-ray sources in Ori\,OB1.  The H$\alpha$ emission 
line sources (CTTSs) are spatially closer to the Orion molecular clouds than the X-ray sources 
(WTTSs), a consequence of CTTSs being less evolved than WTTSs so are still in the proximity of 
their birthplace.

Stars in an OB association share common space motions, so proper-motion measurements can be used to 
identify probable members in an OB association with no limitation on stellar masses.  Because it is 
difficult to measure accurately proper motions of faint stars, this method is restricted to nearby 
OB associations and bright (massive and intermediate-mass) stars \citep{dez99}.  One vulnerability 
of this technique is that the proper motions near the solar apex or antapex are small, so it is 
unsuitable to study regions near these two directions, as is the case for Ori~OB1 which is close to the 
solar antapex.

PMS stars are variable \citep{her94}, so a multi-epoch photometric monitoring can be used to 
identify possible PMS stars in an OB association.  This method ensures that most PMS stars be 
found, but may also pick up other kinds of variables which are difficult to weed out on the basis 
of variability alone.  However, to conduct a multi-epoch survey of a large region is time-consuming.  
Using this method, \citet{bri05} have identified 56 low-mass young stars in the Ori\,OB1a subgroup, 
and 141 in Ori\,OB1b.  Recently, \citet{bri07b} apply a similar method and find nearly 200 low-mass 
PMS stars in the 25 Ori group.  \citet{car01} use the southern 2MASS telescope and perform a 
0.84$\times$6 square degrees multi-epoch photometric observations near Trapezium.  They find 1235 
near-infrared variable stars, most of which ($\sim$93\%) are probably associated with the Orion 
A molecular cloud.  They also show that the spatial distribution of stars with near-infrared excess 
traces the $^{13}$CO emission well.

None of the methods mentioned above are sensitive to those young stars embedded in clouds because of 
excessive extinction.  Recently, {\it Spitzer Space Telescope} is used to investigate disk evolution in 
the $\sigma$ Orionis cluster, 25 Orionis aggregate in Ori OB1a, and Ori OB1b \citep{her07a,her07b}.  
\citet{meg05} conduct a survey in the Orion region with {\it Spitzer} to study star formation 
associated with the A and B molecular clouds.

In this work, we use 2MASS near-infrared data \citep{cut03} to select CTTSs and HAeBe stars, so 
as to diagnose recent star formation related to the Superbubble.  Although near-infrared data 
miss sources deeply inside a cloud, most CTTSs and HAeBe stars can still be detected in the region 
we have investigated.  We describe in \S 2 selection criteria of the young star sample, and follow-up 
spectroscopic observations.  In \S 3, we outline the properties of some specific star-forming 
regions.  We discuss in \S 4 the structure of the Superbubble, and how it influences the star 
formation history out to a distance of some 200~pc.  The conclusions are summarized in \S 5.

\section{OBSERVATIONS AND RESULTS}

\subsection{Selection of PMS Candidate Stars}

Existence of young stellar objects indicates star formation activity in a molecular cloud.  
The time scales (or ages) of protostars are short ($\sim10^{5}$~yr), so the chances of finding protostars 
in a cloud are low even if the cloud has ongoing star formation.  On the other hand, the number of 
PMS stars, with an order of longer evolutionary time scales, should be much higher. 
CTTSs in general are in an earlier evolutionary stage --- hence in closer spatial association 
with molecular clouds --- than the WTTS population, which usually has 
an extended spatial distribution, as seen, e.g., in Taurus \citep{li98}, Chamaeleon \citep{alc97}, 
and Lupus \citep{kra97}.  

It is known that different kinds of PMS stars, namely WTTSs, CTTSs, and HAeBe 
stars occupy different loci in the color-color diagram \citep{lad92}.  
In particular, such a diagram is a useful to tool to identify PMS stars with strong 
infrared excesses, i.e., HAeBe stars and CTTSs, or to distinct young stars from 
classical Be stars \citet{her05}, which are evolved main-sequence early-type stars 
whose excess radiation is attributed mainly to free-free emission, rather than to dust thermal 
emission.  In general, PMS stars associated with reflection nebulae or molecular outflows are redder 
than those without \citep{ken95}.  For example star groups associated with 
Perseus, Orion A, Orion B, and Monoceros R2 molecular clouds   
contains young stellar objects embedded in clouds \citet{car00}.

\citet{lee05} use 2MASS color-color diagram to select CTTS candidates which lie between the two 
parallel lines, $(J-H)-1.7(H-Ks)+0.0976=0$ and $(J-H)-1.7(H-K)+0.450=0$, and above the dereddened 
CTTS locus \citep{mey97}, $(J-H)-0.493(H-Ks)-0.439=0$.  It has been known that HAeBe stars show more 
prominent infrared excesses than CTTSs \citep{lad92,her05}.  Therefore, \citet{lee07} extend the 
selection criteria to include HAeBe stars with colors redder than the line, $(J-H)-1.7(H-Ks)+0.450=0$.  
\citet{lee08} further improve the criteria for HAeBe candidates, resulting the region between 
the two parallel lines, $(J-H)-1.7(H-Ks)+0.450=0$ and $(J-H)-1.7(H-Ks)+1.400=0$, and above the line 
$(J-H)=0.2$, to be more discriminative to exclude contaminating sources such as B[e] and 
classical Be stars from the young star sample.  The improved set of criteria has been used to study 
the star formation in Per~OB1 at high Galactic latitudes \citep{lee08}.

\begin{figure}
\includegraphics[angle=90,scale=1.0]{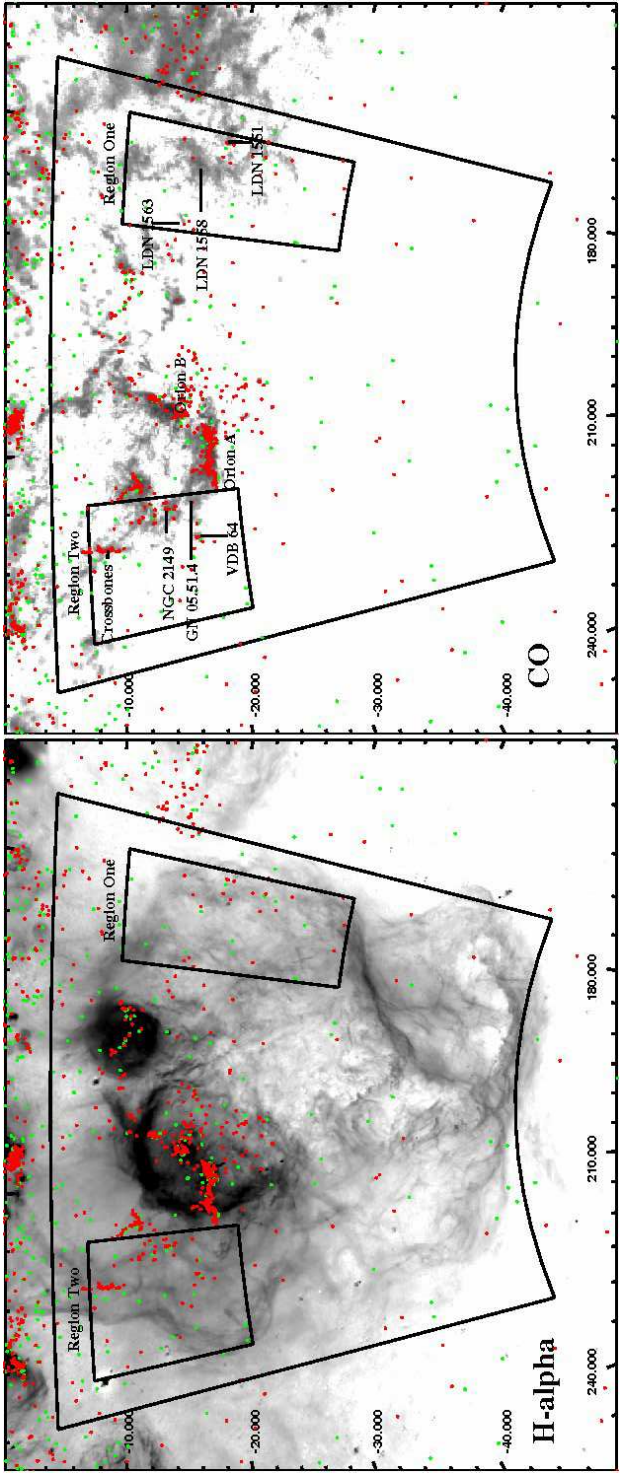}
\caption{H$\alpha$ \citep{fin03} and $^{12}$CO \citep{dam01} images of the Orion-Eridanus 
Superbubble in Galactic coordinates.  The red and green dots are CTTSs and HAeBe candidate stars, 
respectively, which are selected by their near-infrared colors.  Solid lines indicate the region 
studying in this paper.}
\label{fig:superbubble}
\end{figure}

Figure~\ref{fig:superbubble} shows the overview of the Orion-Eridanus Superbubble in H$\alpha$ 
\citep{fin03,den98,gau01,haf03} and CO \citep{dam01} images.  All of the PMS candidate stars which 
were identified by 2MASS colors are plotted.  A total of 692 CTTS and 198 HAeBe candidate stars have 
been identified in the region ($\ell$=175 to 230 and $b= -50$ to $-5$).  It is clear that the PMS 
candidate stars are concentrated in the Orion~A and B giant molecular clouds.  Because these PMS 
candidate stars are grouped or clustered together, they are likely to be PMS stars.  On the other 
hand, some molecular clouds are associated with only a few PMS candidate stars.  These candidates 
need spectroscopic observations to confirm their PMS nature.

\subsection{Spectroscopic Observations}

We took spectroscopic observations of bright PMS candidate stars in two regions, Region One 
($178\degr < l < 188\degr$, $-32\degr < b < -11\degr$) and Region Two ($215\degr < l < 227\degr$, 
$-22\degr < b < -8\degr$).  Region One includes the LDN\,1551, LDN\,1558, and LDN\,1563 molecular 
clouds; Region Two includes NGC\,2149, GN\,05.51.4, VdB\,64, and the Crossbones.  Using the 
criteria described in \S 2.1, there are 21 CTTS and 13 HAeBe candidate stars in Region One, and 
84 CTTS and 25 HAeBe candidates stars in Region Two.

The criteria of \citet{lee07} were used to select PMS star sample in our first observing run at 
Kitt Peak National Observatory.  Spectra of bright PMS stars in Region Two were taken with the 
2.1~m telescope during 2004 January~1--5.  The GoldCamera spectrometer, with a Ford 3K$\times 1$K 
CCD with 15~$\micron$ pixels, was used with the grating \#26new, which gives a dispersion of 
1.24~\AA~pixel$^{-1}$ and covers 3800--7100~\AA.  

In the second observing run, at Beijing Astronomical Observatory, we used the modified criteria 
\citep{lee08} to select our PMS candidate stars.  For these, low-dispersion spectra with a 
dispersion of 200~\AA~mm$^{-1}$ (or 4.8~\AA~pixel$^{-1}$) were taken with the 2.16~m optical 
telescope during 2007 October~19--21.  An OMR (Optomechanics Research, Inc.) spectrograph was 
used with a Princeton Instrument SPEC10 400$\times$1300B CCD detector covering 3500--8500~\AA.  
PMS candidate stars in both regions were included in the second run.  

Spectra of lamp and standard stars were taken during both observing runs, and that would be used 
to perform wavelength and flux calibration.  All the spectroscopic data were processed by the 
standard procedure with the NOAO/IRAF package.  After correcting the bias and flat-fielding, the 
IRAF package KPNOSLIT was used to extract and to calibrate the wavelength and flux of each spectrum.

\subsection{Results}

In the two regions, we obtained spectra for a total of 43 bright PMS star candidates, with 12 
stars in Region One and 31 in Region Two.  Thirty two of these 43 candidates turn out to be 
bona fide PMS stars, with 10 HAeBe and 22 T Tauri stars, giving a success rate of about three quarters 
based on 2MASS colors.   

Checking with SIMBAD\footnote{http://simbad.u-strasbg.fr/simbad/}, we find 21 out of the 32 
spectroscopically confirmed PMS stars new identifications.  Table~\ref{spectra} lists the T Tauri, 
HAeBe stars, and non-PMS sources studied with our spectroscopic observations.  We also determine, 
if applicable, the H$\alpha$, [\ion{O}{1}], and [\ion{S}{2}] equivalent widths of PMS stars.  
T~Tauri stars with H$\alpha$ equivalent widths greater than 10~\AA\,are classified as CTTSs 
\citep{her88}.  In Table~\ref{spectra}, stars 1--11 are located in Region One, and the others are 
in Region Two.  Notably, there are HAeBe stars and CTTSs in Region Two, whereas only CTTSs are 
found in Region One, perhaps because of its smaller PMS sample.  The spatial distribution of PMS 
stars in Table~\ref{spectra} is shown in Figure~\ref{fig:one} and \ref{fig:two}.

Originally , star 22 is classified as a K5 star with low-dispersion spectroscopy \citep{lee05}.  We 
later realize, with higher dispersion and signal-to-noise ratio spectra, that it should be an F6 star.
We also find that intermediate-mass PMS F-type stars, namely stars 22 and 37, do not show the 
lithium absorption line (6708~\AA) in their spectra.

\section{STAR FORMATION}

Using near-infrared data, we can detect CTTSs with strong infrared excesses, but unavoidably  
will miss WTTSs or CTTSs with reduced infrared excesses.  Although a complete census of young stellar 
objects is a crucial first step to study disk fraction evolution or the initial mass 
function, but it is not the purpose of this work.  Instead, our goal is to trace {\it recent\/} star 
formation activity in clouds interacting with the Superbubble.  Therefore, even if our 
PMS star sample may not be complete, it does not affect our results.  Still, we include young stellar 
samples found in the literatures to supplement our sample.

\begin{figure}
\includegraphics[angle=0,scale=0.8]{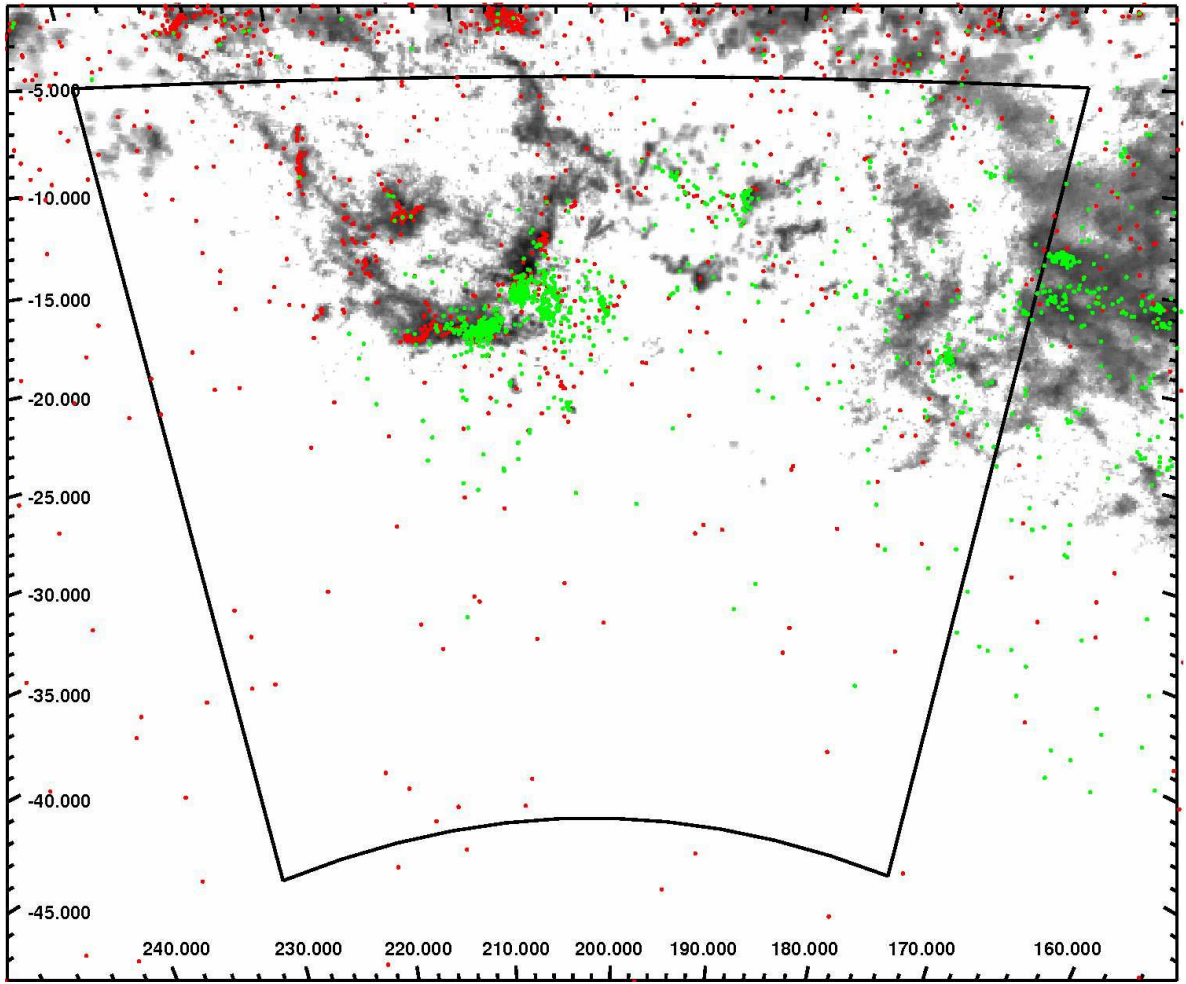}
\caption{Young stellar objects from SIMBAD (green dots) and our sample (red dots) including 
CTTS and HAeBe stars.  SIMBAD objects are distributed over the Orion region and LDN~1551.  However, 
our sample reveal star formation in less explored regions, NGC~2149, VdB~64, and the Crossbones. Solid 
lines indicate the region studying in this paper.}
\label{fig:simbad}
\end{figure}

Searching young stellar objects in SIMBAD including Variable Star of Orion Type (Or*), T 
Tau-type Star (TT*), Young Stellar Object (Y*O), and Variable Star of FU Ori type (FU*), we find 
1647 objects in the same region that we search CTTSs and HAeBe stars from 2MASS.  
Figure~\ref{fig:simbad} shows the distribution of the young stellar objects in SIMBAD.  
Most of the objects are distributed in Ori OB1, and clustered in LDN 1551.  This means that 
previous studies concentrate on these regions.  Compared with our sample, we do not find those 
young stellar objects away form molecular clouds, e.g., in Ori OB1a and 1b.  These objects are 
mainly WTTSs \citep{bri01,bri05,bri07b}, and that is why they are not selected by our criteria.  
We also miss a lot of young stellar objects in LDN 1551.  On the other hand, our sample provides 
PMS star sample in the less explored regions, e.g., NGC~2149, VdB~64, and the Crossbones.  
We cross-correlate our PMS star sample and young stellar objects from SIMBAD for those young 
stars lised in the SIMBAD database for positional coincidence within 5$\arcsec$, and find 86 
CTTS and 20 HAeBe candidate stars in the region.  Hereafter, we call the HAeBe star and CTTS 
candidates as HAeBe stars and CTTSs, respectively, for convenience.

\subsection{Orion A and B Giant Molecular Clouds}

M\,42 is one of the most active nearby star-forming regions.  \citet{hil97} list nearly 1600 
optically detected sources within 2.5~pc of the center.  $Chandra$ detects 1616 X-ray sources 
in the Orion Nebula \citep{get05,fei05}.  Compared to M\,42, NGC\,2024 is more embedded, hence 
it requires X-ray or infrared observations to study the young stars in this region.  $Chandra$ 
detect 283 X-ray sources \citep{ski03}, and \citet{hai01a} find 257 sources in N band 
(10.8~\micron).  M\,78 is less explored than previous two regions.  There are only about six 
sources found in SIMBAD (Figure~\ref{fig:simbad}).

\citet{meg05} use the $Spitzer$ Space Telescope to survey the Orion~A and B molecular clouds.  
Combined with the 2MASS data, they find more than one thousand young stars, approximately half of 
which are grouped in clusters, e.g., NGC~2024 and the Orion Nebula, whereas the other half 
are distributed.  These authors detect 487 sources in M\,42, 239 in NGC\,2024, and 229 in M\,78.  The 
limitation of the study by \citet{meg05} is that they only survey the side of the Orion~A molecular 
cloud with strong $^{13}$CO emission \citep{bal87} and two fields of the Orion~B cloud.  
In comparison, our study based on the 2MASS data does not find as many young stars as by \citet{meg05}, 
but our spatial coverage is much more extended in these two molecular clouds.

\begin{figure}
\includegraphics[angle=270,scale=0.7]{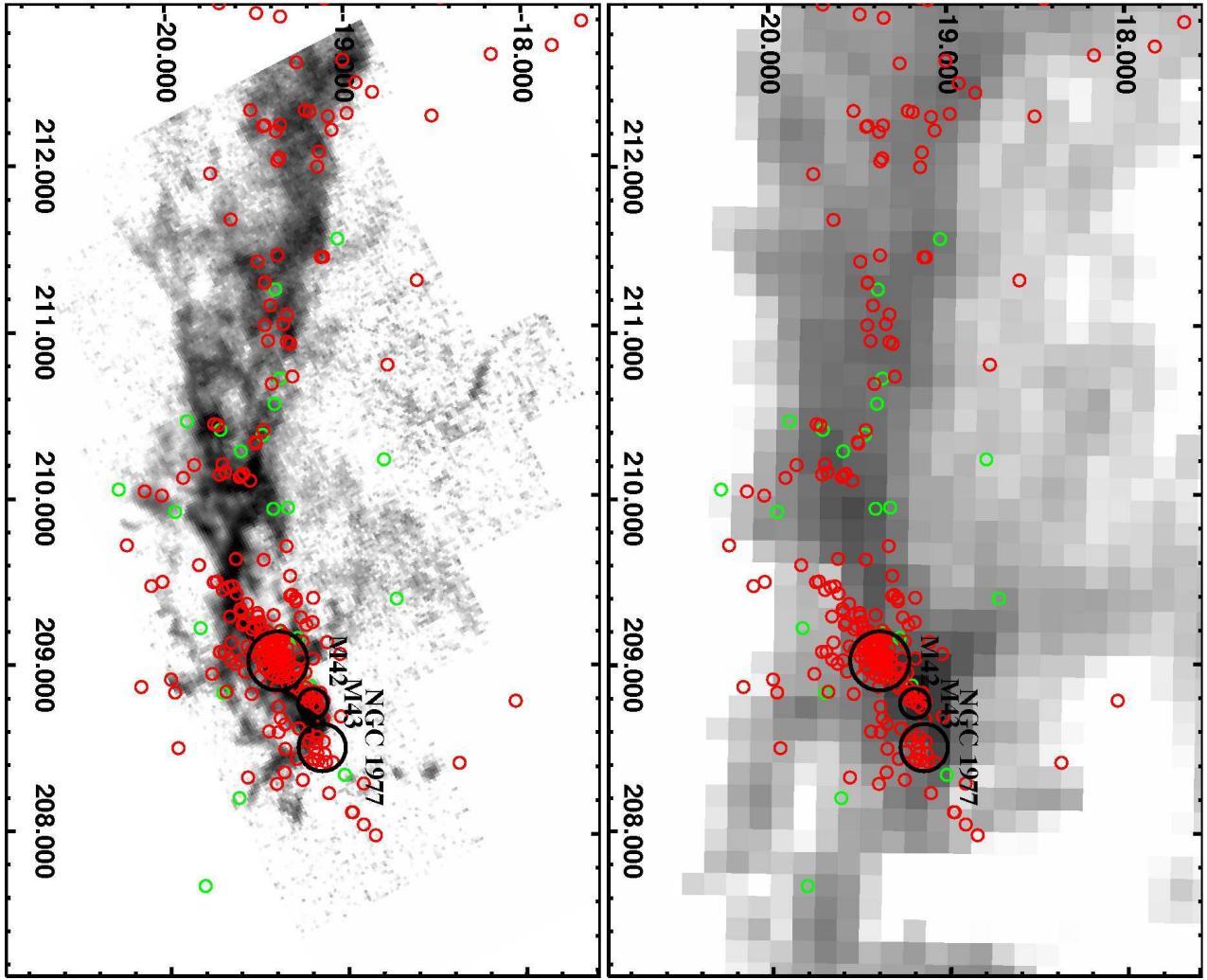}
\caption{$^{12}$CO (top, \citet{dam01}) and $^{13}$CO (bottom, \citet{bal87}) images of the 
Orion~A molecular cloud in Galactic coordinates.  The red and green dots are CTTSs and HAeBe 
candidate stars, respectively, which are selected by their near-infrared colors.}
\label{fig:oria}
\end{figure}

As seen in Figure~\ref{fig:oria}, the intensity of the $^{12}$CO emission of the Orion~A 
molecular cloud is distinctively stronger on one side, like a ridge separating Orion~A into 
two parts.  In Figure~\ref{fig:oria}, the cloud as traced by $^{13}$CO emission shows 
filamentary structure, the most prominent being the integral-shaped filament including M\,43 
and M\,42, which is thought to be the outcome of compression by the Superbubble \citep{bal87}.  
Most PMS stars are associated with $^{13}$CO emission, in particular with more than 
150 PMS stars along the integral-shaped filament.  The level of star-formation activity 
in Orion\,A increases from NGC\,1977, peaks at M\,43 and M\,42, and then drops 
very quickly to the rest of the cloud.

\begin{figure}
\includegraphics[angle=270,scale=0.7]{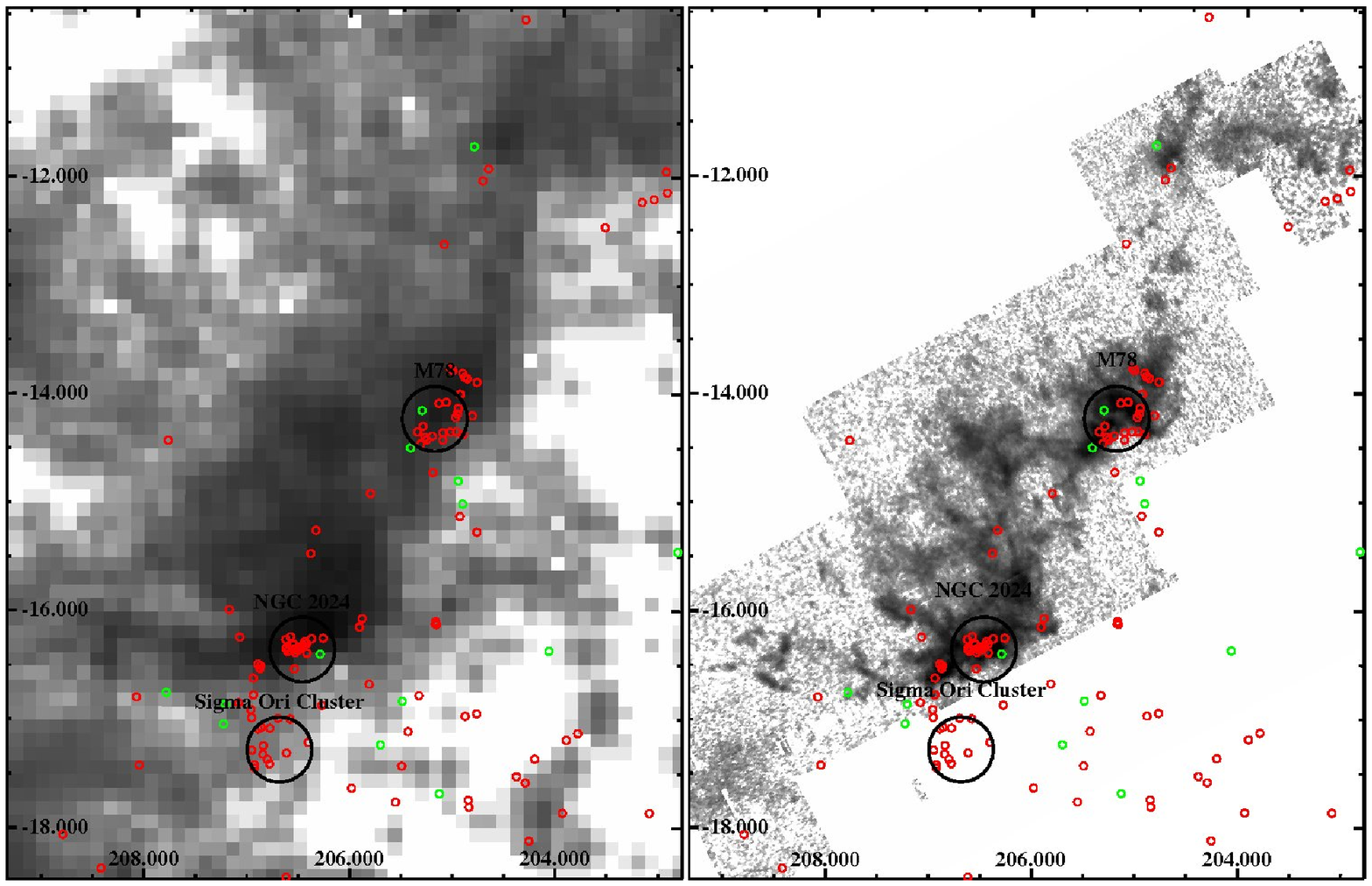}
\caption{Same as Fig.~\ref{fig:oria}, $^{12}$CO (left, \citet{dam01}) and $^{13}$CO (right, 
\citep{mie94}), but for the Orion~B molecular cloud.}
\label{fig:orib}
\end{figure}

The distribution of $^{13}$CO emission in the Orion~B molecular cloud \citep{mie94} is more 
conspicuous on the side facing Ori\,OB1a, where star formation is also more active 
(Figure~\ref{fig:orib}).  Star formation around Orion~B is concentrated in three locations, M\,78 
(including the reflection nebulae M\,78 and NGC\,2071), NGC\,2024, and the $\sigma$~Ori cluster.  
We find some two dozen PMS stars in each of these locations. 

As mentioned above, recent star formation primarily took place on the side facing 
Ori\,OB1a in both the Orion~A and B clouds.  Could there be nondetected CTTS (selected by 
near-infrared colors) embedded on the other side of the clouds?   It is unlikely.  The 
extinction of the regions not associated with CTTSs, as derived from the data in \citet{fro07}, 
is generally low, mostly with A$_{J} \la 0.6$~mag, so any CTTSs, if they exist, could not 
have escaped from the detection by 2MASS.  In comparison, the extinction of star-forming 
clouds in Orion~A and B is typically A$_{J} > 0.8$~mag or much higher.  Our PMS 
sample, naturally limited by the 2MASS sensitivity, thus should be substantially complete 
for CTTSs showing large near-infrared excess.  It is therefore a secured conclusion that 
CTTSs with strong infrared excesses are distributed predominantly on the side facing Ori~OB1a.

\subsection{NGC~2149, GN~05.51.4, VdB~64, and the Crossbones}

Relative to other star-forming regions in Orion\,A and B, NGC\,2149, GN\,05.51.4, VdB\,64, and 
the Crossbones are not as explored.  They have been mapped in $^{12}$CO \citep{wil05,mad86} and 
$^{13}$CO \citep{kim04}.  Figure~\ref{fig:two} depicts the H$\alpha$ \citep{fin03} and extinction 
maps \citep{fro07} of this region.  The CO radial velocities of NGC~2149, GN~05.51.4 and VdB~64 
\citep{wil05,kim04} are similar to that of Ori\,A, with $\sim$~10--12~km s$^{-1}$, so a distance 
of 425~pc is adopted for these clouds.  We find 12 PMS candidates in NGC\,2149, 5 in 
GC\,05\,05.51.4, and 7 in VdB\,64.  Of there, 7 in NGC\,2149, 2 in GN\,05.51.4, and 4 in VdB\,64 
are spectroscopically confirmed to be bona fide T~Tauri stars (Table~\ref{spectra}).

The elongated Crossbones cloud includes the reflection nebula VdB\,80 and the dark cloud LDN~1652.  
There are 42 PMS stars found in the Crossbones (Fig.~\ref{fig:superbubble} and \ref{fig:two}).  
Interestingly, there young stars are not centrally concentrated as in a star cluster, but rather 
are distributed along the elongated cloud.  Stars 39, 40, 42, and 43 in Table~\ref{spectra} and 
stars 20 -- 24 in \citet{lee05} are associated with the Crossbones, the majority (7 out of 9) of 
which exhibit forbidden line(s), indicative of youth \citep{ken98}.  Besides CTTSs, there is an 
HAeBe star, PDS\,23 \citep{vie03,gre92}, in this cloud.

\subsection{Northern Filament}

The Northern Filament is a long, narrow cloud, with an estimated mass of $ 1.7 \times 
10^{4}$~M$_{\sun}$ \citep{wil05}.  It is located near the Orion~B molecular cloud, but closer to 
the Galactic plane (Figure~\ref{fig:superbubble}).  Without knowledge of possible stars 
associated with the cloud, the distance of the Northern Filament can only be inferred by indirect 
methods.  \citet{mad86} estimate a distance of $800\pm 170$~pc from star counts, but adopt a 
distance of 500~pc in view of its similar radial velocity as the nearby Orion\,B cloud, which is 
at a distance of 500~pc.  Recently, \citet{wil05} use foreground and background stars from 
$Hipparcos$ catalogue to determine a distance of 393$^{+65}_{-48}$.  

There is no sign of ongoing star formation in the Northern Filament \citep{mad86}---no \ion{H}{2} 
region or reflection nebula---and indeed we find no apparent PMS star groups in the Northern 
Filament, despite the expectation of some hundreds of young stars in the region \citep{wil05}.  
This cloud thus serves as a good comparison case in an otherwise active star-forming region.

\subsection{LDN~1551, LDN~1558, and LDN~1563}

\begin{figure}
\includegraphics[angle=0,scale=0.7]{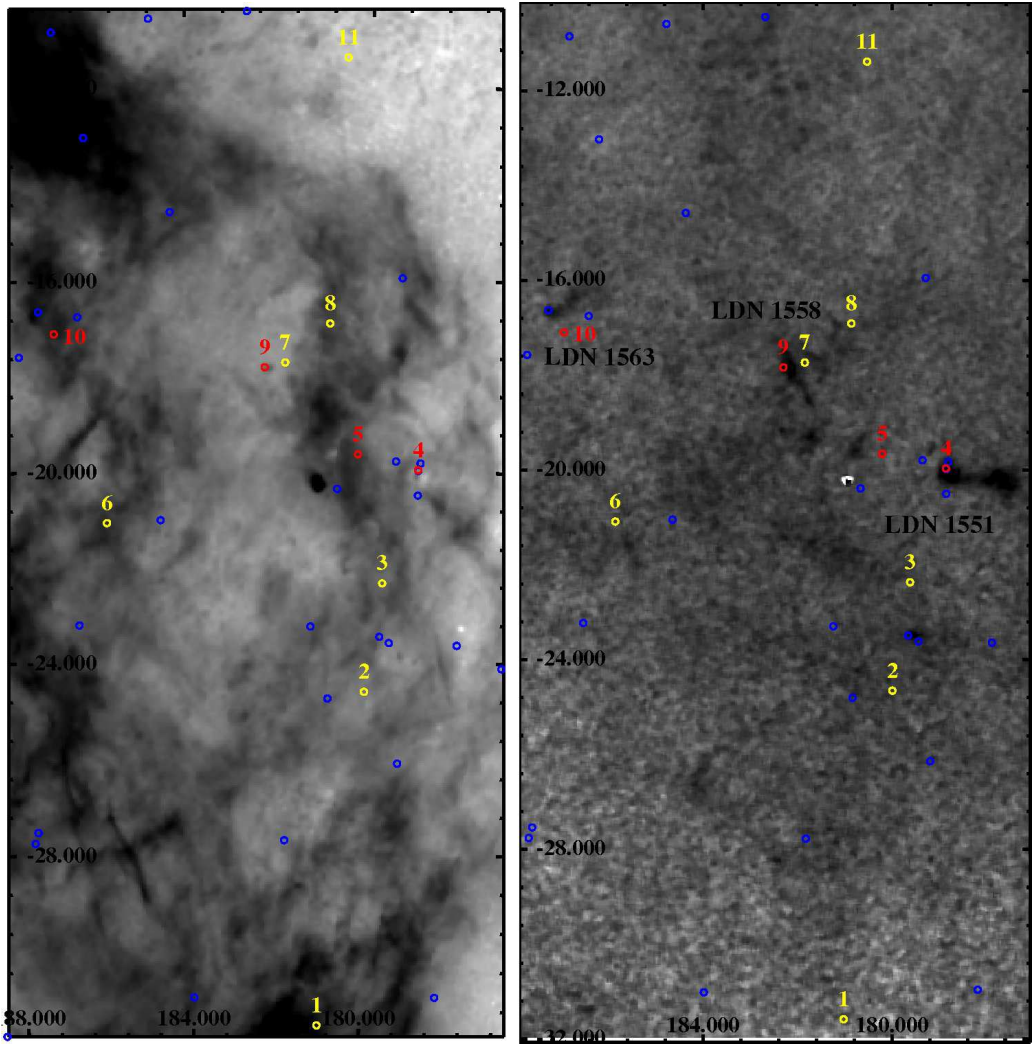}
\caption{H$\alpha$ image (left, \citep{fin03}) and extinction map (right, \citep{fro07}) around the 
Region One.  Stars with spectra type in Table~\ref{spectra} are labeled.  The red, green, yellow, and 
blue circles are CTTSs, HAeBe, non-PMS, and PMS candidate stars selected by 2MASS colors, respectively.}
\label{fig:one}
\end{figure}

\begin{figure}
\includegraphics[angle=90,scale=1.0]{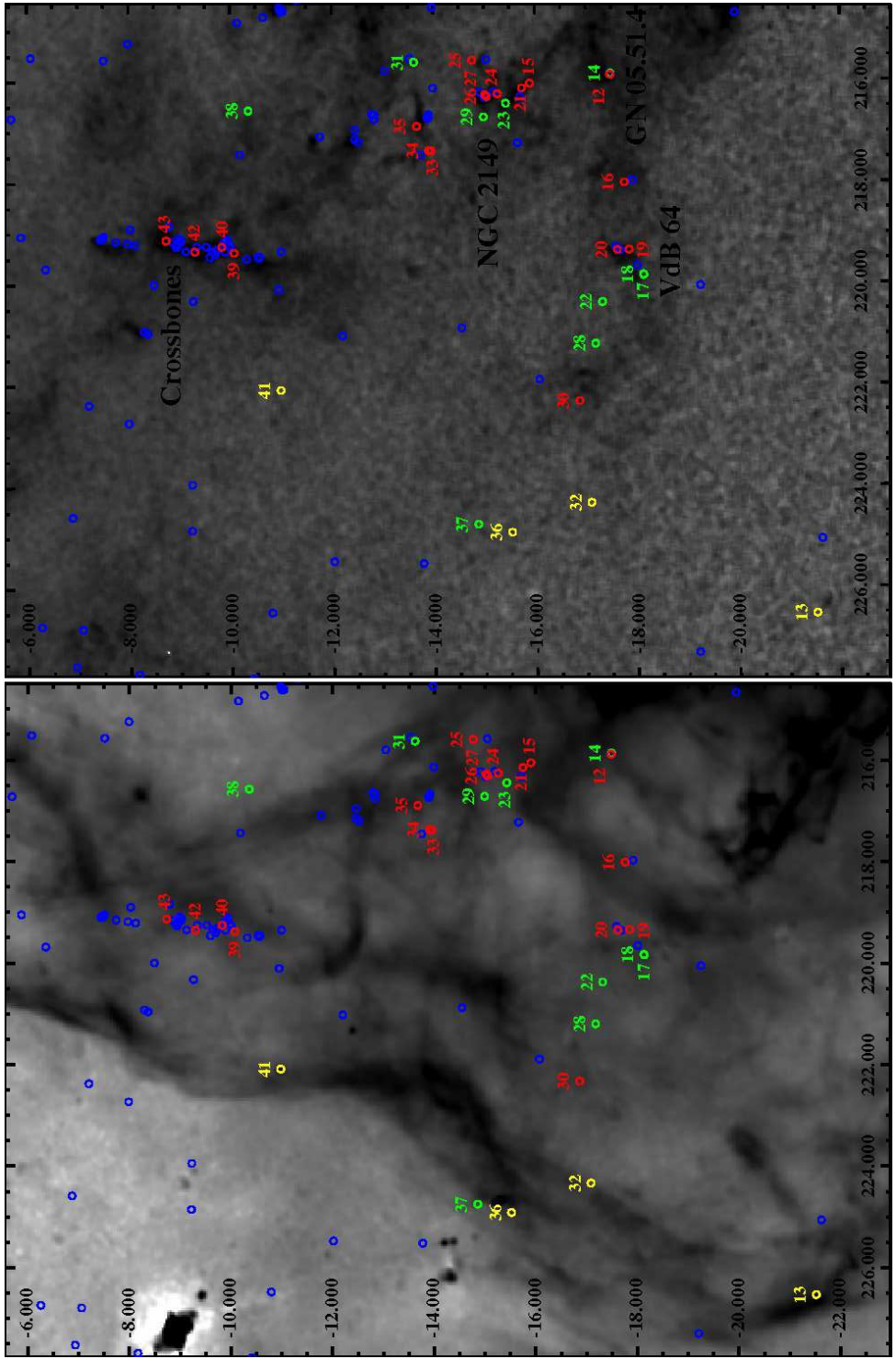}
\caption{Same as Fig.~\ref{fig:one}, but for the Region Two.}
\label{fig:two}
\end{figure}

The clouds LDN\,1551, LDN\,1558, and LDN\,1563 are located on the periphery of the Superbubble, 
outlined by H$\alpha$ filaments.  LDN\,1551 offers one of the best examples of a 
superbubble-cloud interaction.  In Figure~\ref{fig:superbubble} and \ref{fig:one}, the H$\alpha$ 
filament follows the rim of the comet-shaped cloud.  \citet{mor06} find that the cloud points to 
the direction of Orion, and argue that the outline of LDN~1551 is illuminated and eroded by 
Ori\,OB1.

There are at least 30 PMS stars in LDN~1551 \citep{luh06,gom93,bri98,mor06}.  Multi-generational 
star formation in LDN\,1551 is suggested by \citet{mor06}, spanning the past few million years and 
leading to the formation of two small clusters of protostars, together with a halo of more evolved 
PMS stars.  Note that almost all PMS stars are located near the rim of the cloud, whereas protostars 
are embedded inside the cloud.  The distribution of PMS stars is similar to what is seen in the 
bright-rimmed clouds in Ori\,OB1 and Lac\,OB1 \citep{lee05,lee07,che07,che08}, implying sequential 
star formation in LDN~1551.

Star~9 in our list (DR\,Tau), with a distance of 138$^{+82}_{-37}$~pc \citep{ber06}, is likely  
associated with LDN\,1558 (see Fig.~\ref{fig:one}).  Additional young stellar objects around 
LDN\,1558 include the CTTSs, DQ\,Tau and Haro\,6$-$37 \citep{fur05}.  According to the UCAC2 
catalog \citep{zac04}, the proper motion of DQ\,Tau is (4.2$\pm$6.0~mas yr$^{-1}$,$-24.6 \pm 
5.9$~mas yr$^{-1}$), similar to that of DR\,Tau (5.2$\pm$6.2~mas yr$^{-1}$, -26.8$\pm5.4$~mas 
yr$^{-1}$).  On the other hand, the proper motion of Haro\,6$-$37 ($-19.1\pm 6.0$~mas yr$^{-1}$, 
$-57.1 \pm 6.0$~mas yr$^{-1}$) is very different.  We assume that the LDN\,1558 molecular cloud 
is the birthplace for DR\,Tau and DQ\,Tau, so a distance of 138~pc is adopted for LDN\,1558.

LDN\,1563 is associated with Sh\,2$-$246, a nebula with H$\alpha$ emission \citep{fic90}.  
There are no O or early B stars close to LDN\,1563, so the H$\alpha$ emission likely arises from 
the Superbubble.  In addition, the morphology of LDN\,1563 is similar to that of LDN\,1551, both 
appearing comet-shaped pointing to the direction of Ori\,OB1.  Star~10 is projected close to, so 
is likely related to, LDN\,1563.

\section{DISCUSSION}

\subsection{Structure of the Superbubble}

\begin{figure}
\includegraphics[angle=0,scale=0.9]{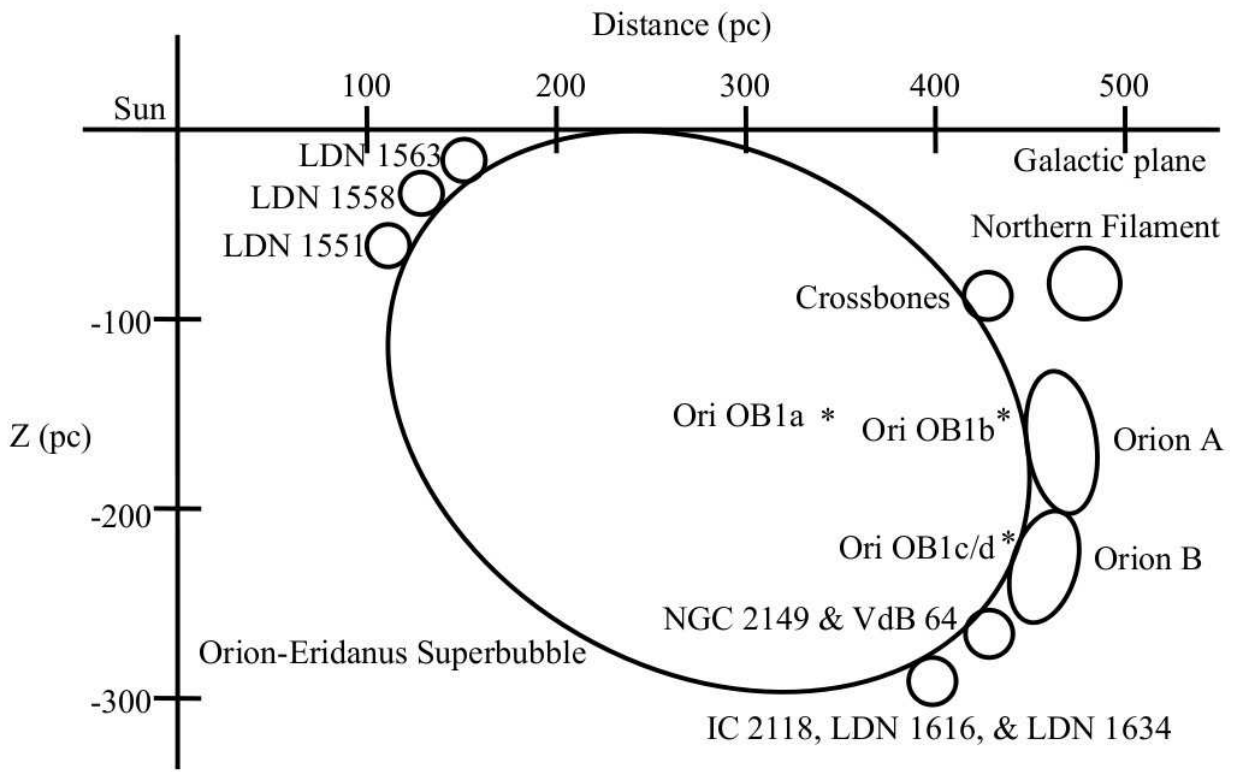}
\caption{Sketch of the Superbubble, approximately to scale.  The shape of the Superbubble, the 
relative positions of clouds, and Ori~OB1 subgroups are ploted.}
\label{fig:sketch}
\end{figure}

It is hard to know the real three-dimensional structure of the Superbubble, because the distance 
to each edge is uncertain.  \citet{bur93} generate a model for the Superbubble, based on what was 
proposed by \citet{rey79}, with the boundary of the Superbubble estimated only circumstantially.  
Since Ori~OB1 is responsible for ionization of the Superbubble \citep{rey79}, those clouds within 
or on the border of the Superbubble should exhibit H$\alpha$ emission.  We sketch the strucure of 
the Superbubble in Figure~\ref{fig:sketch}, in relation with the molecular clouds and the OB 
subgroups in Ori\,OB1.  In addition to the molecular clouds discussed in \S 3, we also include 
here three molecular clouds, IC\,2118 \citep{kun01,kun04}, LDN\,1616 \citep{alc04,gan08}, and 
LDN\,1634 \citep{ste86,dow88}, which have been studied in our previous papers to be outlined also 
by H$\alpha$ filaments \citep{lee05,lee07}.  

Compared with previous works of \citet{rey79} and of \citet{bur93}, our Superbubble model extends 
further to the Galactic plane, as delineated by the H$\alpha$ emission 
(Figure~\ref{fig:superbubble}), with Ori~OB1a located roughly at the center.  Table~\ref{distance} 
lists the distance to each cloud from us, and the distance between the cloud and the center of the 
Superbubble, i.e., Ori~OB1a.  The relative position of each cloud is shown in 
Figure~\ref{fig:sketch}, where one sees a distribution of H$\alpha$ clouds along the border of the 
Superbubble.  This kind of phenomenon has been known in other superbubbles, e.g., the Cepheus ring 
\citep{kun87,bal89}, the Per\,OB1 superbubble \citep{lee08}, the Carina flare \citep{fuk99,daw08}, 
and the \ion{H}{1} loop related to NGC\,281 \citep{sat07}.  

LDN\,1551, LDN\,1558, and LDN\,1563 reside at the near edge of the Superbubble, about 140--180~pc 
away from us.  This is consistent with the 21~cm and Na~D observations by \citet{guo95}, who 
estimate the distance to the near side of the Superbubble to be $ 159 \pm 16$~pc.  As seen in 
Figure~\ref{fig:sketch}, the Northern Filament is located outside the Superbubble.  This explains 
why it is not associated with H$\alpha$ filament or any star formation.

\subsection{Triggered Star Formation}

As discussed in $\S$3, clouds coincident with H$\alpha$ filaments are associated with 
star formation.  A superbubble may play an important role in sequential star formation in 
an OB association \citep[see, e.g., the reviews by][]{ten88,elm98}.  In the Orion~A and 
B molecular clouds that we investigate, PMS stars correlate well with filamentary structures 
($\S$3.1), which were caused by the massive stars in Ori\,OB1 \citep{bal91}.  That young stars 
are closely associated with filaments provides strong evidence of triggered star formation, in that 
massive stars cause clouds to be swept up to high density, i.e., filamentary structures, and then 
stars form in the compressed regions.

\begin{figure}
\includegraphics[angle=0,scale=0.9]{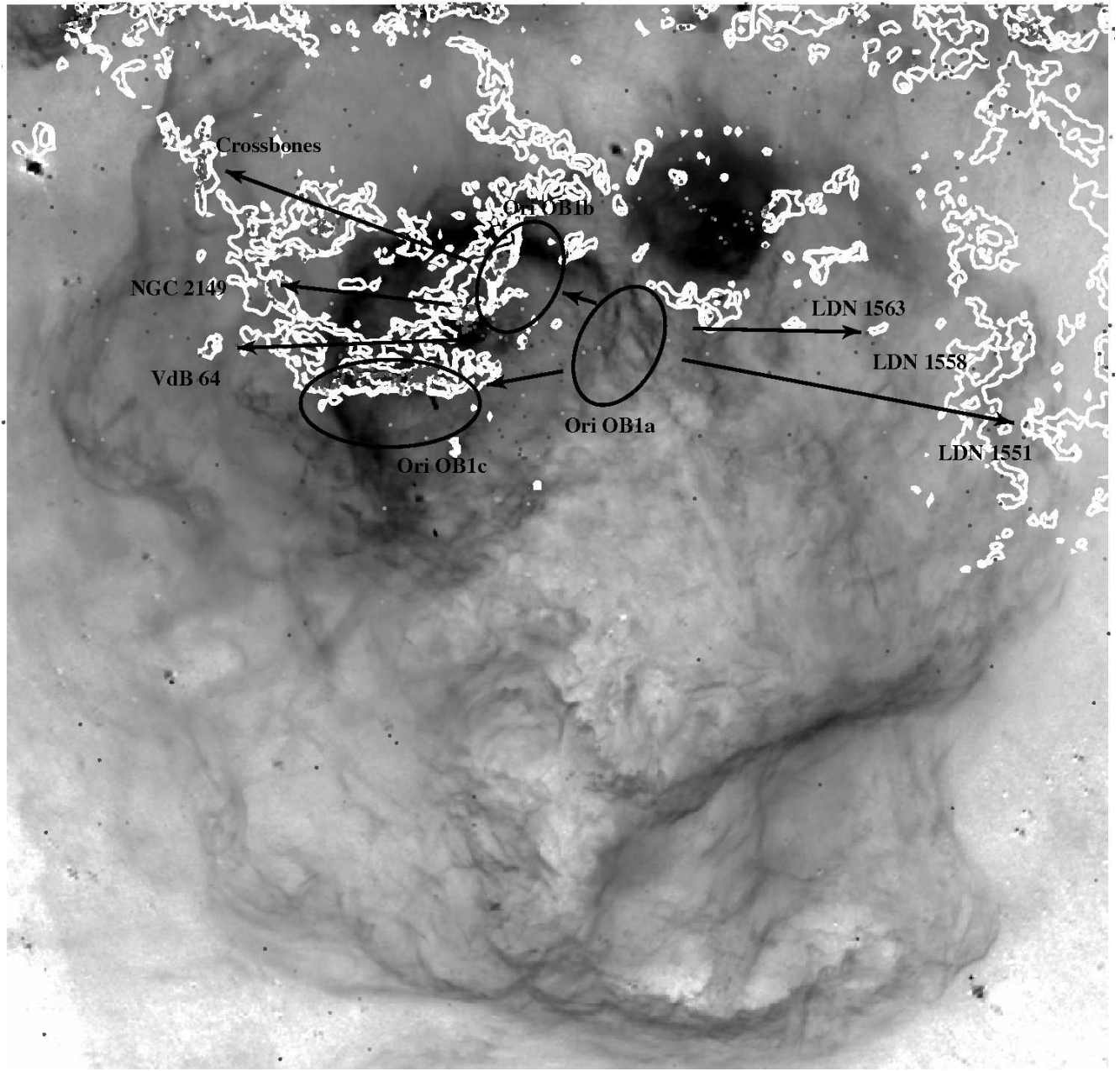}
\caption{Star formation history inside of the Orion-Eridanus Superbubble.  Star formation starts 
from Ori~OB1a and propagates to 1b, 1c, and eventually to 1d.  The Superbubble spreads star formation 
to NGC~2149, VdB~64, and the Crossbones, and furthermore also initiates star formation two hundred pc 
away from Ori~OB1a in the LDN~1551, LDN~1563, and LDN~1558 molecular clouds.  The grayscale and 
contours represent the H$\alpha$ and $^{12}$CO emission, respectively.  The symbols for stars are the 
same as in Figure~\ref{fig:superbubble}.}
\label{fig:SFH}
\end{figure}

\citet{bro94} propose a star formation sequence in the Ori\,OB1 association; namely it started 
from subgroup Ori~OB1a, and propagated to subgroups 1b, 1c, and eventually to 1d.  Our study 
extends the diagnosis of the star formation history to the border of the Superbubble 
(Figure~\ref{fig:SFH}), which was created by the stellar winds and supernova explosions of the 
luminous stars in Ori\,OB1.  The Superbubble expanded and interacted with clouds such as the 
Crossbones, NGC\,2149, GN\,05.51.4, and VdB\,64.  The Crossbones is well outlined by H$\alpha$ 
filaments, which is manifest of interaction (Figure~\ref{fig:two}).  The correlation between 
the compressed high-density gas and existence of young stars in the Crossbones accordingly 
provides a vivid example of star formation triggered by the Superbubble. NGC\,2149, 
GN\,05.51.4 and VdB\,64 are seen also associated with H$\alpha$ filaments (Figure~\ref{fig:two}), 
though not as clear as the Crossbones.  On the opposite side, the Superbubble appears to have 
interacted with and prompted star formation in LDN\,1551, LDN\,1563, and LDN\,1558 as 
well (Figure~\ref{fig:one}).

\citet{bri07a} claim that shock velocities in the range of 15--45~km s$^{-1}$ are able to 
induce star formation in molecular cores.  Velocities faster than 45~km s$^{-1}$ would shred cloud 
cores to pieces whereas velocities slower than 15~km s$^{-1}$ would be unable to trigger a core to 
collapse.  The expansion velocity of the Orion-Eridanus Superbubble is measured to be $\sim$15~km 
s$^{-1}$ \citep{rey79}, marginally high enough to play a constructive role in induced star formation.  
It appears that the Northern Filament provides a case for a cloud neither reaching the critical 
density to collapse spontaneously, nor being compressed by the Superbubble.

Given the distance ($D$) between a cloud and Ori~OB1a (see Table~\ref{distance}), and the expansion 
velocity ($V$)of the Superbubble, the dynamical timescale ($t$) of the Superbubble can be estimated 
by $t \simeq  \eta\, D\, V^{-1}$ \citep{boo08}, where $\eta$ is 0.6 for an energy-conserved bubble 
\citep{cas75,wea77} and 0.4 for a momentum-conserved bubble \citep{ste75}.  Here, we adopt an 
average value $\eta$=0.5 to derive a dynamical timescale $t$ to be in the range of 4.3--8.8~Myr for 
the near and far edges of the Superbubble.

In a triggered star-formation scenario, the age of an O star, if it is still in existence, must be 
at least comparable to the ages of next-generation young stars plus the shock traveling time of a 
superbubble.  
In our case, the Superbubble was created by the massive stars in Ori\,OB1a, whose 
main-sequence lifetimes are some 11~Myr \citep{bro94}.  
Because a typical CTTS would lose its disk in 6~Myr 
\citep{hai01b}, the CTTSs that we have found, all with ample NIR excesses, should be no more than a 
few million years old.  Therefore, the causality holds to support a triggered star formation process 
in terms of the ages of Ori~OB1a (11.4~Myr), the transverse time of Superbubble (4.3--8.8~Myr), and 
typical ages of CTTSs ($ \la 6$ Myr).

\section{CONCLUSIONS}

We have conducted a census of classical T Tauri and Herbig Ae/Be stars selected by 2MASS colors in the 
vicinity of the Orion-Eridanus Superbubble produced by Ori~OB1.  Our study results in young star 
samples in two giant molecular clouds 
(Orion~A and B) and additional smaller molecular clouds, including NGC~2149, GN~05.51.4, VdB~64, the 
Crossbones, the Northern Filament, LDN~1551, LDN~1558, and LDN~1563.  Combined with H$\alpha$ and CO 
images, we investigate the star formation history in this region.  The main results of our work are 
summarized as follows:

\begin{enumerate}

\item Clouds associated with H$\alpha$ filaments are preferentially located on the border of the 
Orion-Eridanus Superbubble, created by UV photons from luminous stars in Ori~OB1.  Clouds outlined by 
H$\alpha$ filaments exhibit recent star formation.  

\item The Northern Filament neither accompanies H$\alpha$ filament, nor is associated with PMS stars.  
It is likely to be situated outside, hence out of reach of the compression by, the Superbubble to form 
stars.

\item Star formation triggered by the expanding Orion-Eridanus Superbubble started from Ori~OB1a and 
propagated to other subgroups.  The Superbubble not only triggered starbirth in Orion~A and B, but 
also led to current star formation in NGC~2149, GN~05.51.4, VdB~64, and the Crossbones which are 
located some one hundred pc away from Ori~OB1a.

\item The Superbubble likely also prompted star formation in clouds some two hundred pc away from 
Ori~OB1a, as witnessed in LDN\,1551, LDN\,1563, and LDN\,1558.

\item The triggered star formation scenario is supported by sequential star formation, ages of PMS stars 
in the clouds, and the expansion timescale of the Superbubble.

\end{enumerate}

\acknowledgments
We thank the anonymous referee for detailed reading, constructive comments, and suggestions.  We 
want to thank specially Richard F. Green who kindly provided us KPNO director's discretionary time 
for this study.  Our gratitude goes to John Bally for providing $^{13}$CO maps of Ori~A and Ori~B 
and to You-Hua Chu for helpful discussions.  We thank Jinzeng Li and the staff at the Beijing 
Astronomical Observatory for their assistance during our observing runs.  This research makes use 
of data from the Two Micron All Sky Survey (2MASS) and the Southern H-Alpha Sky Survey 
Atlas (SHASSA).  We acknowledge the financial support of the grants NSC95-2119-M-008-028 and 
NSC95-2745-M-008-002 of the National Science Council of Taiwan.

\clearpage


\begin{deluxetable}{cclccl}
\tablecaption{Spectral Observations}
\tablewidth{0pt}
\tabletypesize{\footnotesize}
\rotate
\tablehead{\colhead{Star} & \colhead{2MASS} & \colhead{Emission Line(s)\tablenotemark{a}} & \colhead{Li\tablenotemark{b}} & \colhead{Type} & \colhead{Remarks} \\}
\startdata
01 & J03595596+0919044 &                                        & - & C         &                                             \\ 
02 & J04191911+1427140 &                                        & - & C         &                                             \\ 
03 & J04254487+1610141 &                                        & - & C         & C* 3243                                     \\ 
04 & J04314007+1813571 & H(-71.5), O(-3.7), S(-3.7), Ca, Fe, He & - & CTTS      & XZ Tau, associated with LDN~1551            \\ 
05 & J04360131+1726120 & H(-30.8)                               & - & CTTS\tablenotemark{d} &                                             \\ 
06 & J04442877+1221118 &                                        & - & AGN       & LEDA 2816425                                \\ 
07 & J04462180+1723031 &                                        & - & C         & V1027 Tau                                   \\ 
08 & J04464456+1837505 &                                        & - & M         &                                             \\ 
09 & J04470620+1658428 & H(-66.7), Ca, Fe, He                   & - & CTTS      & DR Tau, associated with LDN~1558            \\ 
10 & J05001726+1348576 & H(-37.7), O(-4.0), Ca                  & - & CTTS\tablenotemark{d} & close to LDN~1563                    \\ 
11 & J05045222+2210340 &                                        & - & C         & IRAS 05018+2206                             \\ 
12 & J05534559-1024510 & H(-203.3), Ca                          & - & CTTS\tablenotemark{d} & associated with GN~05.51.4                                            \\ 
13 & J05550798-2113049 &                                        & - & AGN       & z=0.285                                     \\ 
14 & J05534254-1024006 & H(-43.4), O(-1.3), S(-0.3)             & n & HAeBe(B7) & IRAS 05513-1024, associated with GN~05.51.4 \\ 
15 & J05595090-0952488 & H(-145.2), O(-1.7), Fe, He, Na         & a & CTTS\tablenotemark{d} & IRAS F05574-0952                            \\ 
16 & J05561496-1221599 & H(-6.0)                                & - & WTTS\tablenotemark{d} &                                             \\ 
17 & J05574918-1406080 & H\tablenotemark{c}                     & - & HAeBe\tablenotemark{d} & associated with VdB~64                                            \\ 
18 & J05574947-1405336 & H\tablenotemark{c}                     & - & HAeBe     & IRAS 05555-1405, associated with VdB~64                                            \\ 
19 & J05580493-1332592 & H(-33.3)                               & - & CTTS\tablenotemark{d} &                                             \\ 
20 & J05585792-1327412 & H(-33.1)                               & - & CTTS\tablenotemark{d} &                                             \\ 
21 & J06003451-0953341 & H(-114.6)                              & - & CTTS\tablenotemark{d} &                                             \\ 
22 & J06014515-1413337 & H(-4.4)                                & n & HAeBe(F6) & IRAS 05594-1413, (LCZ2005) 13                            \\ 
23 & J06021488-1000595 & H(-75.1), Fe                           & n & HAeBe(B4) & V791 Mon, IRAS 05598-1000                   \\ 
24 & J06023181-0946510 & H(-22.8)                               & n & CTTS\tablenotemark{d} &                                             \\ 
25 & J06031315-0859167 & H(-15.3)                               & - & CTTS\tablenotemark{d} &                                             \\ 
26 & J06032324-0944118 & H(-357)                                & - & CTTS\tablenotemark{d} & associated with NGC~2149                                            \\ 
27 & J06032727-0941271 & H(-100.4), Ca                          & - & CTTS\tablenotemark{d} & associated with NGC~2149                                            \\ 
28 & J06033705-1453025 & H(-15.8)                               & n & HAeBe(B6) & AE Lep, IRAS 06013-1452                     \\ 
29 & J06041782-1003349 & H(-8.5)                                & n & HAeBe(B7)\tablenotemark{d} & IRAS 06019-1003                             \\ 
30 & J06063868-1544195 & H(-16.7)                               & - & CTTS\tablenotemark{d} & close to GN 06.03.6                                            \\ 
31 & J06072583-0831044 & H(-7.3)                                & n & HAeBe(A9)\tablenotemark{d} & IRAS 06050-0830                             \\ 
32 & J06090348-1735089 &                                        & n & AGN       & z=0.104                                     \\ 
33 & J06091346-1011386 & H(-22.1)                               & - & CTTS\tablenotemark{d} &                                             \\ 
34 & J06092371-1009549 & H(-31.5)                               & - & CTTS\tablenotemark{d} &                                             \\ 
35 & J06092550-0938492 & H(-142.9), O(-3.3), Fe                 & a & CTTS      & AS 119, IRAS 06070-0938                     \\ 
36 & J06160006-1727270 &                                        & n & C         & (LCZ2005) 29                                            \\ 
37 & J06181637-1702347 & H\tablenotemark{c}                     & n & HAeBe(F8) & UY CMa                                      \\ 
38 & J06205595-0754362 & H(-21.3)                               & n & HAeBe(A4)\tablenotemark{d} &                                             \\ 
39 & J06265390-1015349 & H(-48.3), O(-3.8), He, Fe              & n & CTTS      & NSV 2968, associated with Crossbones        \\ 
40 & J06273428-1002397 & H(-1.8)                                & a & WTTS      & (LCZ2005) 23, associated with Crossbones                  \\ 
41 & J06281742-1303109 & H(-179.3), O(-2.7), Fe                 & n & B[e]      & with [\ion{Fe}{2}] line emission, HD 45677  \\ 
42 & J06294017-0953463 & H(-6.2),O(-0.6)                        & a & CTTS\tablenotemark{d} & IRAS 06272-0951, associated with Crossbones \\ 
43 & J06312038-0927047 & H(-131.1), O(-4.3), Fe, He             & a & CTTS\tablenotemark{d} & IRAS 06289-0924, associated with Crossbones \\ 

\enddata
\tablenotetext{a}{H--H$\alpha$, O--[\ion{O}{1}], S--[\ion{S}{2}], He--\ion{He}{1}(6678), Fe--Fe, Ca--\ion{Ca}{2}, Na--\ion{Na}{1}, The number following H, O, S, are the equivalent widths of H$\alpha$, [\ion{O}{1}], and [\ion{S}{2}], respectively.}
\tablenotetext{b}{- -- low resolution spectrum; n -- no detection; a -- absorption}
\tablenotetext{c}{Hydrogen Balmer lines are in absorption but filled in by H$\alpha$ emission.}
\tablenotetext{d}{New identified PMS star.}
\label{spectra}
\end{deluxetable} 


\begin{deluxetable}{lllll}
\tablecaption{Distances of Clouds}
\tablewidth{0pt}
\tabletypesize{\footnotesize}
\rotate
\tablehead{\colhead{Region} & \colhead{Heliocentric distance} & \colhead{Distance to Ori~OB1a} & \colhead{Reference} & \colhead{Star Formation} \\
                            & \colhead{(pc)}     & \colhead{(pc)} &                         &        }
\startdata
LDN~1551             & 139       & 212 & \citet{ber99}       & T Tau stars only            \\
LDN~1558             & 138       & 209 & \citet{ber06}       & T Tau stars only            \\
LDN~1563             & 180       & 264 & \citet{cla91}       & T Tau stars only            \\
NGC~2149             & 425       & 134 & \citet{wil05}       & HAeBe stars and CTTSs       \\
GN~05.51.4           & 425\tablenotemark{a} & 129 & \citet{wil05,kim04} & HAeBe star and CTTS   \\
VdB~64               & 425       & 148 & \citet{wil05}       & HAeBe stars and CTTSs       \\
Crossbones           & 457       & 184 & \citet{wil05}       & HAeBe stars and CTTSs       \\

\enddata
\tablenotetext{a}{The CO radial velocity of GN~05.51.4 \citep{kim04} is similar to NGC~2149, so we assume that the distances of GN~05.51.4 and NGC~2149 are the same.}
\label{distance}
\end{deluxetable}

\clearpage

\clearpage

\clearpage

\clearpage

\clearpage

\clearpage

\clearpage

\clearpage

\end{document}